\documentclass[notitlepage,twocolumn,superscriptaddress,amsmath,amssymb,aps,prl]{revtex4-1}
\pdfoutput=1
\usepackage[utf8]{inputenc}
\usepackage[T1]{fontenc}
\usepackage{hyperref}
\usepackage{graphicx}
\usepackage{amsfonts, amsmath, amsthm, amssymb} 
\usepackage{braket}
\usepackage{float}
\usepackage{amsthm}
\usepackage{color}
\usepackage[normalem]{ulem}
\usepackage[all]{nowidow}
\usepackage[dvipsnames]{xcolor} 
\usepackage{tabularx}
\usepackage{physics} 

\hypersetup{citecolor=black,colorlinks=false,urlcolor=black} 
\usepackage[english]{babel}

\usepackage{pdfpages}
\makeatletter
\AtBeginDocument{\let\LS@rot\@undefined}
\makeatother

\usepackage[pdftitle={Mapping  Phase Diagrams of Quantum Spin Systems through Semidefinite-Programming Relaxations}]{}
\usepackage{soul}
\usepackage{tikz}
\usepackage{bbold}

\begin{document}

\title{Mapping  Phase Diagrams of Quantum Spin Systems through \\ Semidefinite-Programming Relaxations}

\author{David Jansen}
\affiliation{ICFO-Institut  de  Ciències  Fotòniques,  The  Barcelona  Institute  of  Science  and  Technology, 08860 Castelldefels (Barcelona), Spain}

\author{Donato Farina}
\affiliation{Physics Department E. Pancini - Università degli Studi di Napoli Federico II, Complesso Universitario Monte S. Angelo - Via Cintia - I-80126 Napoli, Italy}
\affiliation{Istituto Nazionale di Fisica Nucleare (INFN), Sezione di Napoli, Complesso Universitario Monte S. Angelo, Via Cintia, 21, Napoli, 80126, Italy.}

\author{Luke Mortimer}
\affiliation{ICFO-Institut  de  Ciències  Fotòniques,  The  Barcelona  Institute  of  Science  and  Technology, 08860 Castelldefels (Barcelona), Spain}
\author{Timothy Heightman}
\affiliation{ICFO-Institut  de  Ciències  Fotòniques,  The  Barcelona  Institute  of  Science  and  Technology, 08860 Castelldefels (Barcelona), Spain}
\affiliation{Quside Technologies SL, Carrer d’Esteve Terradas, 1, 08860 Castelldefels, Barcelona, Spain}
\author{Andreas Leitherer}
\affiliation{ICFO-Institut  de  Ciències  Fotòniques,  The  Barcelona  Institute  of  Science  and  Technology, 08860 Castelldefels (Barcelona), Spain}

\author{Pere Mujal}
\affiliation{ICFO-Institut  de  Ciències  Fotòniques,  The  Barcelona  Institute  of  Science  and  Technology, 08860 Castelldefels (Barcelona), Spain}

\author{Jie Wang}
\affiliation{State Key Laboratory of Mathematical Sciences, Academy of Mathematics and Systems Science, Chinese Academy of Sciences, Beijing, China}

\author{Antonio Ac\'{i}n}
\affiliation{ICFO-Institut  de  Ciències  Fotòniques,  The  Barcelona  Institute  of  Science  and  Technology, 08860 Castelldefels (Barcelona), Spain}
\affiliation{ICREA-Institució Catalana de Recerca i Estudis Avan\c cats, Lluís Companys 23, 08010 Barcelona, Spain}

\date{\today}

\begin{abstract}
\noindent
Identifying quantum phase transitions poses a significant challenge in condensed matter physics, as this requires methods that both provide accurate results and scale well with system size. In this work, we demonstrate how relaxation methods can be used to generate the phase diagram for one- and two-dimensional quantum systems. To do so, we formulate a relaxed version of the ground-state problem as a semidefinite program, which we can solve efficiently. Then, by taking the resulting vector of moments for different model parameters, we identify all phase transitions based on their cosine similarity. Furthermore, we show how spontaneous symmetry breaking is naturally captured by bounding the corresponding observable. Using these methods, we reproduce the phase transitions for the one-dimensional transverse field Ising model and the two-dimensional frustrated bilayer Heisenberg model. We also illustrate how the phase diagram of the latter changes when a next-nearest-neighbor interaction is introduced. 
Overall, our work demonstrates how relaxation methods provide a novel framework for studying and understanding quantum phase transitions.
\end{abstract}
\maketitle
{\it Introduction.---} 
\begin{figure}[t]
\includegraphics[width=0.5\textwidth]{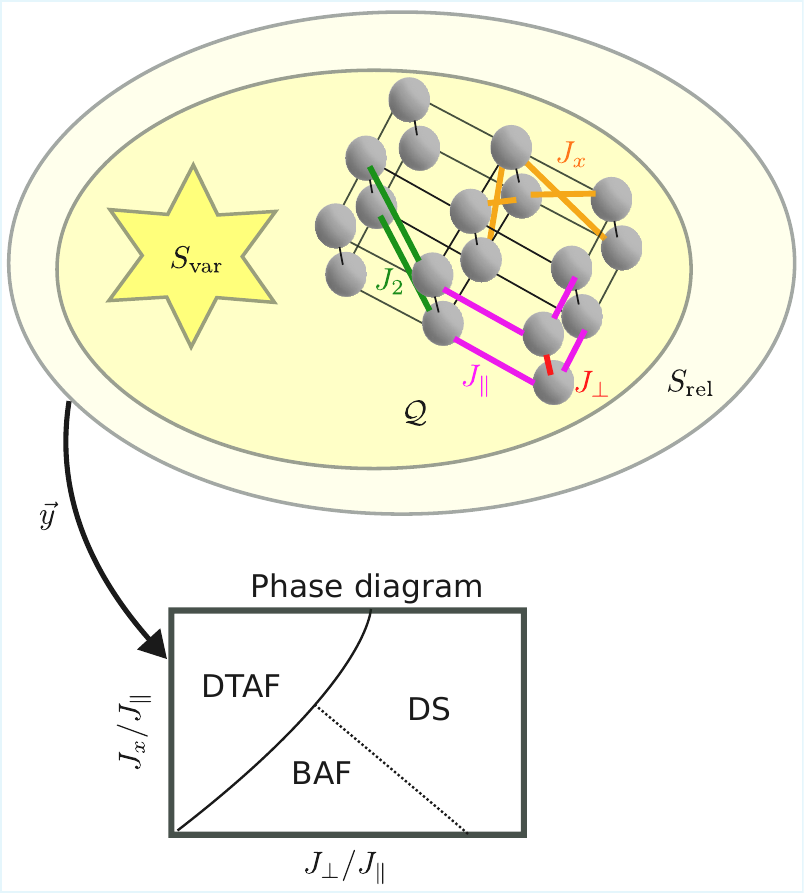}
  \caption{Illustration of the scheme presented in this work. We study a quantum spin system [here we draw the FBH-NNN model, see Eq.~\eqref{eq:hamFBHNNN}] whose moments form a convex set $\mathcal{Q}$. A (possibly) nonconvex set of moments is captured by a variational ansatz, $S_{\textrm{var}} \subseteq \mathcal{Q} $, and the set obtained from relaxation, $S_{\textrm{rel}}$, includes $\mathcal{Q}$, i.e.,  $\mathcal{Q}\subseteq S_{\textrm{rel}}$. From the SDP, we extract the moment vector $\vec{y}$, which we use to map out the phase diagram (here sketched for $J_2/J_{\parallel}=0$ with first- and second-order transitions illustrated by solid and dashed lines, respectively).}
  \label{fig:sketch}
\end{figure}
When and how quantum systems undergo phase transitions is one of the most prominent research questions in condensed matter physics. At zero temperature, for example, one can drastically change the physical properties of a system by tuning parameters in the Hamiltonian. A system could go from being conducting to insulating, or from being in an ordered ferromagnetic phase to being disordered~\cite{Sachdev_2011}. In the context of quantum spin systems, we see a variety of different phases and transitions depending on the nature of the interaction and the lattice geometry~\cite{yan_11,corboz_13,savary_17,Stapmanns_18,bishop_19,gallegos_25,corboz_25}. To study such phenomena, we are faced with another prominent problem:
finding and analyzing the ground state and its properties. 

Over the last decades, numerical methods have become a cornerstone in helping us better understand such quantum phase transitions, see, e.g., Refs.~\cite{sandvik_10,yan_11,corboz_13,Stapmanns_18,sprague2024variational,sprague2024variational,gallegos_25,corboz_25,schuyler_25}. 
In particular, two approaches have gained prominence: exact diagonalization~\cite{sandvik_10} and variational methods (e.g., tensor networks~\cite{white_92,orus_14} and neural network quantum states~\cite{carleo_17}). Exact methods provide the correct values, but due to the exponential scaling of the Hilbert space with the number of lattice sites, studies are limited to relatively small systems. Variational methods, on the other hand, have proven to be scalable to large systems or even to work directly in the thermodynamic limit~\cite{vidal_07,jordan_08,Crosswhite_08,Zauner-Stauber_18}. Despite their success, some drawbacks remain: i) They only provide upper bounds for the true ground-state energy (see Ref.~\cite{wu_24} for examples of models that are difficult for current state-of-the-art variational methods). ii) Determining the quality of the bound can be difficult~\cite{wu_24}. iii) They sometimes optimize over nonconvex landscapes and thus are prone to getting stuck in local minima. iv) They provide no guarantees on the accuracy of the expectation value of a given observable. 

Relaxation methods aim to complement the two aforementioned approaches. Here, the idea is to solve an easier (or relaxed) version of the problem, but in a way that  both provides certified bounds and is more scalable than exact methods. 
The difference between variational and the relaxation method used in this work can be understood as follows:  denote by $\mathcal{Q}$ the set of all moments, e.g., $\{ \expval{\hat \sigma^x_i},\expval{\hat \sigma^z_i},\hdots \}$ for a spin system, for which there exists a quantum state $\hat \rho$ such that $\expval{\hat \sigma^x_i}=Tr[\hat \sigma^x_i \hat \rho]$. A variational ansatz is restricted to a (possibly nonconvex) subspace, $S_{\textrm{var}}$, with $S_{\textrm{var}} \subseteq \mathcal{Q} $, as is illustrated in Fig.~\ref{fig:sketch}. The relaxed problem on the other hand, is defined over a set, $S_{\textrm{rel}}$, such that $\mathcal{Q}\subseteq S_{\textrm{rel}}$. Thus, if one minimizes, e.g., the energy, over $S_{\textrm{rel}}$ ($S_{\textrm{var}}$), one is guaranteed to find an energy that is smaller (larger) than or equal to that of the true ground state. We also want to emphasize that, for the relaxations considered here,  $S_{\textrm{rel}}$ is convex, and therefore the optimization is not affected by local minima. The key challenge, however, is making the bounds as tight as possible. 

Even though the idea of relaxation itself is not new in many-body physics~\cite{baumgratz_2012,barhel_12,mazziotti_04,mazziotti_11}, these methods have recently regained attention. Novel applications have included bounding energies and expectation values of ground states~\cite{kull_24,wang_24}, of steady states~\cite{robichon_24,mortimer_24}, at finite temperatures~\cite{fawzi_24}, during time evolution~\cite{lawrence_24,arujo_25}, and bounding spectral gaps~\cite{rai_24}. 

In this Letter, we show that, by relaxing the ground-state problem to a hierarchy of semidefinite programs (SDPs)~\cite{nevascues_07,nevascues_08,pironio_10} (also see Refs.~\cite{vandenberghe_96,Skrzypczyk_23} for introductions to SDPs), we can efficiently map out the phase diagram of quantum spin systems. We validate our method on the one-dimensional transverse field Ising (TFI) model and the two-dimensional frustrated bilayer Heisenberg model (FBH), later extending it to include next-nearest-neighbor interactions (FBH-NNN). By extracting the moment vectors from the SDP, we infer the phase diagram in an unsupervised fashion using only the cosine similarity as a metric. The input to the algorithm is only the Hamiltonian and physically motivated moments (the scheme introduced here is illustrated in Fig.~\ref{fig:sketch}). We also find that selected moments approximate the true physical behavior of the system well and that spontaneous symmetry breaking is reflected in the bounds on observables.

The rest of this Letter is organized as follows: we first introduce the models and methods and then demonstrate how to map out a complete phase diagram on the three aforementioned systems. We then conclude and discuss further research directions enabled by our method.

{\it Models and method.---}
We begin by briefly describing the known phases of the TFI and FBH models, all in natural units and with periodic boundary conditions. The (average) energy of the $L$-site one-dimensional TFI model~\cite{Sachdev_2011} is given by
\begin{equation}
\label{eq:tfi}
\begin{split}
    E_{\textrm{TFI}}&=\sum\limits_{j=1}^{L} \left(-J\expval*{\sigma^{z}_{j}\hat \sigma^{z}_{j+1}}+h\expval*{\hat \sigma^{x}_{j}}\right) \, .
    \end{split}
\end{equation} In the thermodynamic limit, this model has a second-order phase transition at $h/J=1$, from a degenerate ordered ferromagnet for $h/J<1$ to unordered phase for $h/J>1$ (here, we only consider $J>0$). 
 
Next, recall the FBH model, whose energy reads as
\begin{equation}
\label{eq:fbh}
\begin{split}
    E_{\textrm{FBH}}&=\frac{1}{4}\sum\limits_{\alpha \in \{ x,y,z\}}\sum\limits_{i,j=1}^{L}\Big( J_{\perp} \expval*{\hat \sigma^{\alpha}_{1,i,j}\hat \sigma^{\alpha}_{2,i,j}}+\\
    & \sum\limits_{a=1}^{2}\big( J_{\parallel}(\expval*{\hat \sigma^{\alpha}_{a,i,j}\hat\sigma^{\alpha}_{a,i+1,j}}+\expval*{\hat\sigma^{\alpha}_{a,i,j} \hat\sigma^{\alpha}_{a,i,j+1}}) \\
&+J_{x}(\expval{\hat \sigma^{\alpha}_{a,i,j}\hat \sigma^{\alpha}_{\bar{a},i+1,j}}+\expval{\hat \sigma^{\alpha}_{a,i,j}\hat \sigma^{\alpha}_{\bar{a},i,j+1}}) \big)\Big) \, ,
    \end{split}
\end{equation} where $\bar{a}$ denotes the layer opposite of $a$, $J_{\perp}$ and $J_x$ are nearest and next-nearest neighbor interlayer couplings and $J_{\parallel}$ is the nearest neighbor intralayer coupling (see Fig.~\ref{fig:sketch}). This model has a dimer singlet (DS), a dimer-triplet antiferromagnet (DTAF), and a bilayer antiferromagnet (BAF) phase with one second-order and two first-order transitions between these phases. We note that this model poses significant numerical challenges, since there is a negative sign problem in parts of the phase diagram, making them inaccessible to Quantum Monte Carlo~\cite{Stapmanns_18}. Using infinite projected entangled pair states, the phase diagram was mapped out in Ref.~\cite{Stapmanns_18}.

Finally, we introduce next-nearest neighbor (NNN) interlayer interaction to the FBH model, \begin{equation}
\label{eq:hamFBHNNN}
E_{\textrm{FBH-NNN}}=E_{\textrm{FBH}}+E_{\textrm{NNN}} \, ,
\end{equation}
where
 \begin{equation}
\label{eq:NNN}
\begin{split}
    E_{\textrm{NNN}}=\frac{J_{2}}{4}&\sum\limits_{\alpha \in \{ x,y,z\}}\sum\limits_{i,j=1}^{L}\sum\limits_{a=1}^{2} \Big( \expval*{\hat \sigma^{\alpha}_{a,i,j}  \hat \sigma^{\alpha}_{a,i+1,j+1}}\\  &+\expval*{\hat \sigma^{\alpha}_{a,i,j}\hat \sigma^{\alpha}_{a,i+1,j-1}}\Big) \, .
    \end{split}
\end{equation}
To our knowledge, the phase diagram of this model has not yet been studied. 

Having introduced these three models, we now illustrate our method. First, we reformulate the ground-state problem as a polynomial minimization problem over noncommuting variables. We follow Ref.~\cite{wang_24} and define the following SDP: 
\begin{equation}
\label{sdp-notation}
\begin{aligned}
& \underset{\vec{y} \in \mathbb{R}^m}{\text{min}}\quad \vec{b} \cdot \vec{ y} \\
& \text{\quad s.t.} \quad M=C-\sum\limits_{i=1}^m y_i A_i  \succcurlyeq 0   \,.\\
\end{aligned}
\end{equation}
Here, the moment matrix $M_{\vec \alpha, \vec \beta}:=\langle \hat P_{\vec \alpha} \hat P_{\vec \beta} \rangle $, 
with the ${\hat P}_{\vec \alpha}$'s representing a subset of multiqubit Pauli operators, e.g.,
\begin{eqnarray}
\label{eq:P}
    {\hat P}_{\vec \alpha}={\hat \sigma}^{\alpha_1}_1\otimes\dots{\hat \sigma}^{\alpha_L}_L, 
    \quad {\hat \sigma}^{\alpha_j}_j \in \{ \mathbb{1}, {\hat \sigma}_x, {\hat \sigma}_y, {\hat \sigma}_z \}\,,
    \label{multiqubit-pauli}
\end{eqnarray}
for a one-dimensional chain. Furthermore, $\vec{y}$ is a vector of moments, e.g., $\vec{y} \,{}^T=(1, \expval{\hat \sigma^x_{1}}, \expval{\hat \sigma^y_{1}}, \expval{\hat \sigma^z_{1}}, \expval{\hat \sigma^x_{1}\hat \sigma^y_{2}}, \hdots)$, and $y_i$ is its $i$-th component. Here, we define $b \in \mathbb{R}^m$ such that $\vec{b} \cdot  \vec{y}$ gives the desired energy in terms of moments. Furthermore, $A_i$ and $C$ are ad-hoc Hermitian matrices useful for rewriting the moment matrix, incorporating the Pauli replacement rules \cite{pauli-replacement} and a series of symmetries of the Hamiltonian (see Supplementary Material~\cite{suppmat}). $M\succcurlyeq 0$ means that the matrix $M$ is positive semidefinite.

The constraint $M\succcurlyeq 0$ is a relaxation of state positivity $\hat \rho\succcurlyeq 0 $, meaning that $\hat \rho\succcurlyeq 0 \implies M\succcurlyeq 0$ \cite{nevascues_07, nevascues_08}.
As such, by enforcing only $M\succcurlyeq 0$ we obtain a lower bound on the ground-state energy because we minimize on a set $S_{\rm rel}$ that contains $\mathcal{Q}$ (see Fig.~\ref{fig:sketch}).
We remark that by keeping the number of variables of the problem (and hence the size of the moment matrix) and the number of constraints $\sim poly(L)$, the SDP, Eq.~\eqref{sdp-notation}, comes as a scalable relaxation of the problem of finding the exact ground-state energy~\cite{sdp-eff}.   

As shown in Ref.~\cite{wang_24}, we can also bound expectation values by solving the following SDP:
\begin{equation}
\label{sdp-obs-notation}
\begin{aligned}
& \underset{\vec{y} \in \mathbb{R}^m}{\text{min/max}}\quad \vec{o} \cdot \vec{y} \\
& \text{\quad s.t.} \quad M=C-\sum\limits_{i=1}^m y_i A_i  \succcurlyeq 0   \,,\\
& \qquad \qquad \vec{b} \cdot \vec{y} \leq E_{\textrm{var}}\,,\\
& \qquad \qquad \vec{b} \cdot \vec{y} \geq E_{\textrm{rel}}\,,\\
\end{aligned}
\end{equation}
where $\text{min/max}$ means that we either maximize or minimize depending on whether we want a lower or upper bound, respectively.  The scalars $\vec{o} \cdot \vec{y}$ and $\vec{b} \cdot \vec{y}$ give the expectation values of the desired observable and Hamiltonian expressed as moments respectively, and $E_{\textrm{rel}}$ and $E_{\textrm{var}}$ are the ground-state energies obtained by solving Eq.~\eqref{sdp-notation} and using a variational ansatz. For the TFI model we use $L=30$ sites, and for FBH and FBH-NNN models we use $L=6$. All technical details of the SDP, including moments used, symmetries exploited, etc., can be found in Ref.~\cite{suppmat}.

To identify the phase transitions without any prior knowledge of the system beyond the SDP inputs, we take inspiration from Refs.~\cite{Canabarro_19,kottmann_20,kottmann_21}. There, the authors identified phase transitions in an unsupervised fashion.
Our approach consists of computing the cosine similarity between the moment vectors $\vec y_i$ and $\vec y_j$,

\begin{equation}
\label{eq:sc}
    S_C(\vec{y}_i,\vec{y}_j)=\frac{\vec{y}_i \cdot \vec{y}_j}{ || \vec{y}_i ||\,|| \vec{y}_j || } \, .
\end{equation}
These vectors are obtained from the solutions (i.e., the \textit{argmin}) of the optimization problem defined in Eq.~\eqref{sdp-notation} at two given points in the phase diagram.
The motivation for this choice comes from the relationship between the cosine similarity and the fidelity (see the End Matter).

Apart from mapping out the phase diagram, another key result is that we show that some moment values {from the \textit{argmin} of} Eq.~\eqref{sdp-notation} qualitatively reflect the behavior of the true physical state. 
Note that this is the convex-optimization analog of the approach used with variational methods, where a low-energy state is found, and its properties are assigned to the ground state.

{\it Results.---}
\begin{figure}
  \includegraphics[width=0.5\textwidth]{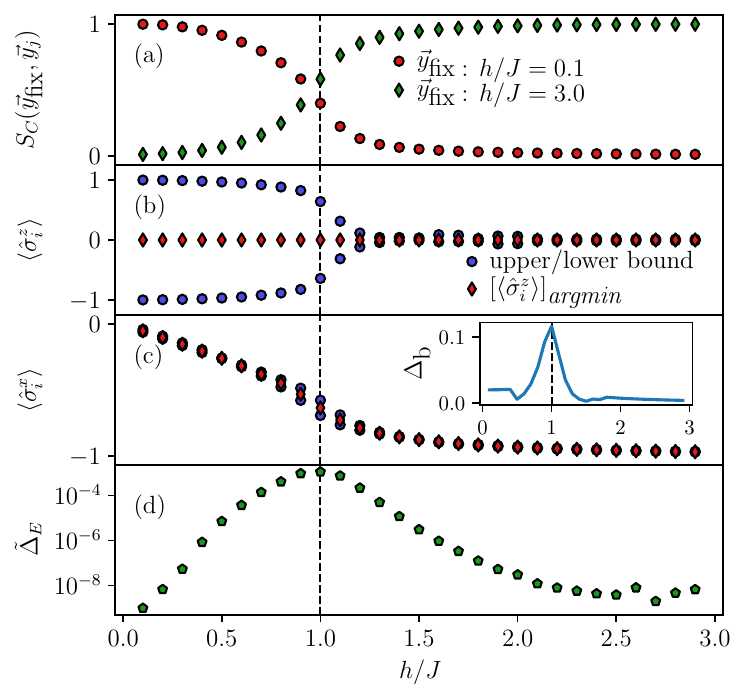}
    \caption{Different quantities extracted from the SDP for the TFI model as a function of $h/J$. The dashed line indicates the transition point at $h/J=1.0$. (a) Cosine similarity of the two vectors $\vec{y}_{\textrm{fix}}$ and $\vec{y}_j$. For the circles, we fix $\vec{y}_{\textrm{fix}}$ to $h/J=0.1$ and iterate $\vec{y}_j$ over all different values of $h/J$. For the diamonds, we follow a similar procedure but fix $\vec{y}_{\textrm{fix}}$ to $h/J=3.0$. 
    (b) Values of the monomial $\expval*{\hat \sigma^z_i}$. The red diamonds are the monomial values that we obtain by solving the minimization problem [Eq.~\eqref{sdp-notation}], and the blue circles are bounds that we get by solving Eq.~\eqref{sdp-obs-notation}.
    (c) Similar to (b) but for $\expval*{\hat \sigma^x_i}$. Inset: difference in the bounds $\Delta_{\textrm{b}}$ [see Eq.~\eqref{eq:bounddiff}].
    (d) Relative difference between the upper and lower bounds for the ground-state energy [see Eq.~\eqref{eq:bounddiff_E}]. 
    }
    \label{fig:tfi}
\end{figure}
\begin{figure*}
  \includegraphics[width=\textwidth]{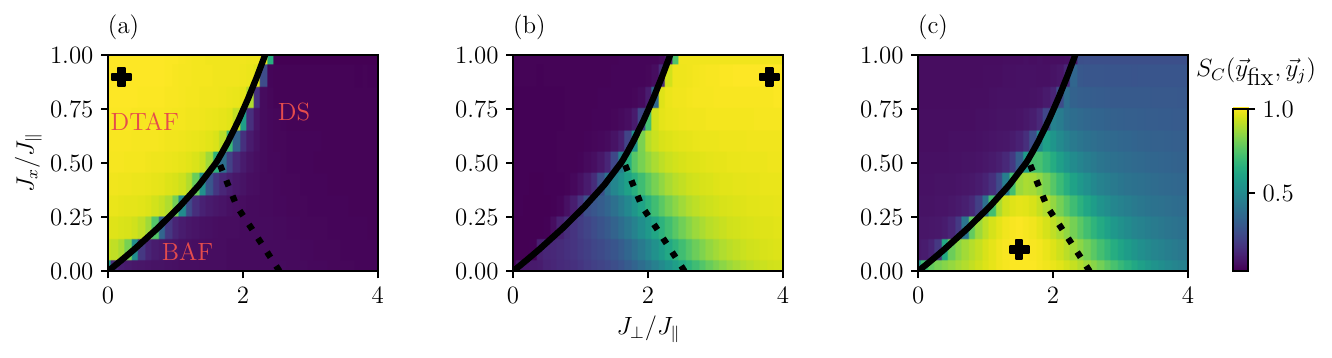}
  \caption{Phase diagram for the FBH model calculated using $S_C(\vec{y}_{\textrm{fix}}, \vec{y}_{j})$. The black solid and dotted lines show the first- and second-order transitions from Ref.~\cite{Stapmanns_18}, respectively. The black crosses show the points that we use for $\vec{y}_{\textrm{fix}}$. Those points are $(J_{\perp}/J_{\parallel},J_x/J_{\parallel} )=(0.2,0.9) $ in (a), $(3.8,0.9)$ in (b), and $(1.5,0.1)$ in (c).}
  \label{fig:phase_diagram}
\end{figure*}
First, we show results for the TFI model. In Fig.~\ref{fig:tfi}(a), we plot $S_C(\vec{y}_{\textrm{fix}},\vec{y}_j)$, where $\vec{y}_{\textrm{fix}}$ is the vector of moments at one point in the phase diagram and $\vec{y}_j$ iterates over all other parameter values. There, one sees how $S_C(\vec{y}_{\textrm{fix}},\vec{y}_j)$ is large within the same phase and rapidly decays across the phase transition. Figures~\ref{fig:tfi}(b) and (c) show the value from the moment vector together with the upper and lower bounds from Eq.~\eqref{sdp-obs-notation}. Here, we see that the moment value is similar to that of the true ground state, which is also often observed in variational calculations. However, this is not guaranteed in either.  The blue circles on the other hand are certified bounds which cannot be obtained via variational calculations. For $\expval{\hat \sigma^z_i}$, the system exhibits a spontaneous symmetry breaking for $h/J<1$, as the energies of the states $\ket{\uparrow\uparrow\uparrow\hdots }$ and $\ket{\downarrow\downarrow\downarrow\hdots }$ become degenerate. This is reflected in the bounds, but not in the moment value. In the inset of Fig.~\ref{fig:tfi}(c), we also show the difference in bounds,
\begin{equation}
\label{eq:bounddiff}
    \Delta_{\textrm{b}}=\expval{\hat \sigma^x_i}_{\textrm{up}}-\expval{\hat \sigma^x_i}_{\textrm{low}} \, ,
\end{equation} which peaks at the point of the transition. We note an important difference between Figs.~\ref{fig:tfi}(b) and \ref{fig:tfi}(c). Theoretically, one should be able to systematically improve the upper and lower bounds in Fig.~\ref{fig:tfi}(c), making it increasingly difficult to detect the phase transition based on the difference between the bounds (though this might not happen in practice). In Fig.~\ref{fig:tfi}(b) on the other hand, $\Delta_{\textrm{b}}$ will always increase to $2$, due to the spontaneous symmetry breaking.

\begin{figure*}[t]
  \includegraphics[width=\textwidth]{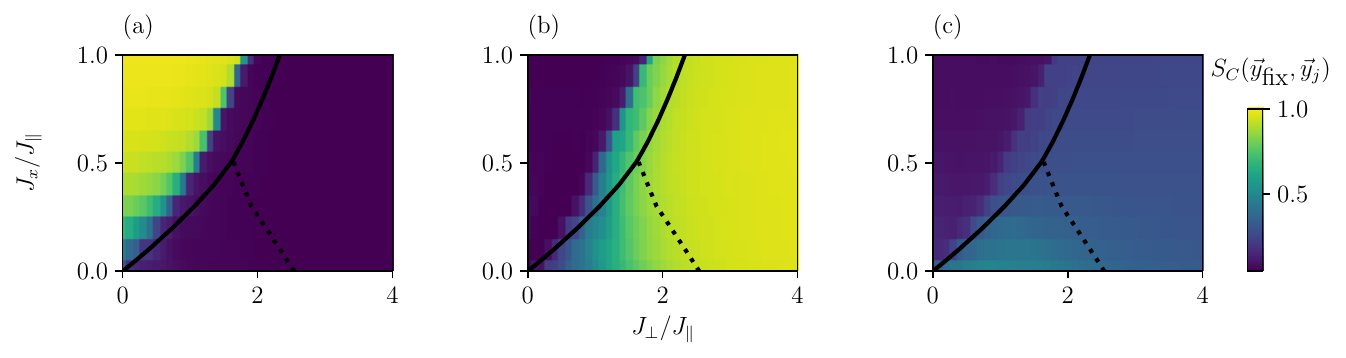}
  \caption{Phase diagram for the FBH-NNN model calculated using $S_C(\vec{y}_{\textrm{fix}}, \vec{y}_{j})$. In (a)-(c), we use $\vec{y}_{\textrm{fix}}$ from Figs.~\ref{fig:phase_diagram}(a)-\ref{fig:phase_diagram}(c) respectively (i.e., $J_{2}/J_{\parallel}=0$). The black solid and dotted lines show the first- and second-order transitions for $J_{2}/J_{\parallel}=0$ and are meant to illustrate the differences. }
  \label{fig:phase_diagram_J2}
\end{figure*}

Figure~\ref{fig:tfi}(d) shows the relative difference between the upper and the lower energy bounds used for the constraints in Eq.~\eqref{sdp-obs-notation}, \begin{equation}
\label{eq:bounddiff_E}
    \tilde{\Delta}_{E}=\frac{E_{\textrm{up}}-E_{\textrm{low}}}{\abs{E_{\textrm{up}}}} \, .
\end{equation} There, we see that the relative difference between the bounds is small, even though it increases around the transition point.

We now demonstrate that this scheme can resolve the complete phase diagram of the significantly more involved FBH model. Even though the phase diagram is well understood, we describe how the procedure would work as though it were unknown. We do this in order to validate the unsupervised nature of our framework. For that, our approach is as follows: We first compute $S_C(\vec{y}_{\textrm{fix}}, \vec{y}_j)$ for one $\vec{y}_{\textrm{fix}}$ and $ \vec{y}_j$ being all points in parameters space. The choice for $ \vec{y}_{\textrm{fix}}$ should preferably be deep in one phase and can, for example, be chosen based on certain known limits of the model or a grid search. Here we choose $(J_{\perp}/J_{\parallel},J_x/J_{\parallel} )=( 0.2,0.9)$ for $\vec{y}_{\textrm{fix}}$, and the results are shown in Fig.~\ref{fig:phase_diagram}(a). There, a transition (which is known to be of first order~\cite{Stapmanns_18}) is clearly visible. Then, by visually inspecting this result, we choose our next point far away from the transition line where $S_C(\vec{y}_{\textrm{fix}}, \vec{y}_j)$ is small (now, we choose $(J_{\perp}/J_{\parallel},J_x/J_{\parallel} )=( 3.8,0.9)$ for $\vec{y}_{\textrm{fix}}$). Again, the first-order transition line is detected [see Fig.~\ref{fig:phase_diagram}(b)], but additionally, we identify a new region of interest (the BAF phase) where $S_C(\vec{y}_{\textrm{fix}}, \vec{y}_j)$ slowly decreases.  This gives us our third choice for $\vec{y}_{\textrm{fix}}$, $(J_{\perp}/J_{\parallel},J_x/J_{\parallel} )=( 1.5,0.1)$, and the suspicion of a third phase is confirmed [see Fig.~\ref{fig:phase_diagram}(c)]. In total, we have identified the three different phases of the model. Whereas many properties of the systems can be inferred by inspecting the moment vector, accurately characterizing the different regions as the DTAF, BAF, and DS phases would require further studies. Furthermore, even though the first-and second-order transitions are visually distinguishable in our data, the order of the transition is not guaranteed to be captured correctly (which, e.g.,  can be seen by the fact that the singlet density in the moment vector is continuous in Fig.~\ref{fig:Jx06} in the End Matter). Thus, complementing the cosine similarity approach with variational calculations and our SDP bounds on observables is necessary for properly characterizing the transition.

Now, we apply our method to the FBH-NNN model with $J_{2}/J_{\parallel}=1/2$. To do that, we proceed as follows. We take the three previously used vectors $\vec{y}_{\textrm{fix}}$ from the $J_{2}/J_{\parallel}=0$ data and compute the overlap with all $\vec{y}_j$ for $J_{2}/J_{\parallel}=1/2$ [Figs.~\ref{fig:phase_diagram_J2}(a)-\ref{fig:phase_diagram_J2}(c)]. We observe that a phase quantitatively similar to the DTAF phase remains ($\max (S_C(\vec{y}_{\textrm{fix}}, \vec{y}_{j}))= 0.97$) but that the first-order transition is shifted to smaller $J_{\perp}/J_{\parallel}$ for large $J_x/J_{\parallel}$ [Fig.~\ref{fig:phase_diagram_J2}(a)]. In Fig.~\ref{fig:phase_diagram_J2}(b), we see that a phase with substantial overlap with the DS phase also survives ($\max (S_C(\vec{y}_{\textrm{fix}}, \vec{y}_{j}))= 0.96$), but in Fig.~\ref{fig:phase_diagram_J2}(c), we see that there is only limited overlap ($\max (S_C(\vec{y}_{\textrm{fix}}, \vec{y}_{j}))= 0.48$) with the BAF phase. This indicates that we might be close to a transition point as a function of $J_{2}/J_{\parallel}$. This is confirmed in Fig.~\ref{fig:multiple_phases} in the End Matter.  

{\it Conclusion.---} In this work, we demonstrated how relaxation methods can be used to efficiently map out the phase diagram of quantum spin systems. With our approach, we correctly identified all the phase transitions of the one-dimensional TFI model and the FBH model. We also found significant changes in the phase diagram for the FBH-NNN model with $J_{2}/J_{\parallel}=1/2$ compared to when $J_{2}/J_{\parallel}=0$.
    
Furthermore, we showed that the moments can reproduce the behavior of physical observables and that, by bounding expectation values of observables, we can also identify phase transitions, particularly when a symmetry is spontaneously broken.
In total, our work extends the application of relaxation methods to systems with unprecedented complexity and allows for efficient detection of phase transitions. 

There are many intriguing extensions of our work. For example, one could bound observables for systems that are difficult for variational methods~\cite{wu_24}. It would also be interesting to test our approach on systems with topological phases~\cite{arujo_21}. Furthermore, the tight bounds achieved in Ref.~\cite{wang_24} inspire our choice of monomials in the SDP, as detailed in~\cite{suppmat}. However, a more efficient selection (see, e.g., Ref.~\cite{requena_23}) might improve our results.

 Our code and data can be found at~\cite{gitrepos_1,gitrepos_2,data}.

\section{Acknowledgments}
This project has received funding from the European Union’s Horizon
2020 research and innovation program under the Marie Sklodowska-
Curie Grant Agreement No 847517, PNRR MUR Project No. PE0000023-
NQSTI, University of Catania via PNRR-MUR Starting Grant Project
PE0000023-NQSTI, European Union (PASQuanS2.1, 101113690, and
Quantera Veriqtas and Compute), MICIN and Generalitat de Catalunya
with funding from the European Union, National Key R\&D Program
of China under Grant No. 2023YFA1009401, the Government of Spain
(Severo Ochoa CEX2019-000910-S, FUNQIP, and European Union NextGen-
erationEU PRTR-C17.I1), Fundació Cellex, Fundació Mir-Puig, Gen-
eralitat de Catalunya (CERCA program), the ERC AdG CERQUTE
and the AXA Chair in Quantum Information Science.
 
 \textit{Data
availability} —The data that support the findings of this article are openly available~\cite{gitrepos_1,gitrepos_2,data}.
 \section{  
 {End Matter}
 }
 \label{sec:em}
  \subsection{  
 Further details on the frustrated bilayer Heisenberg model
 }
\begin{figure}
  \includegraphics[width=0.5\textwidth]{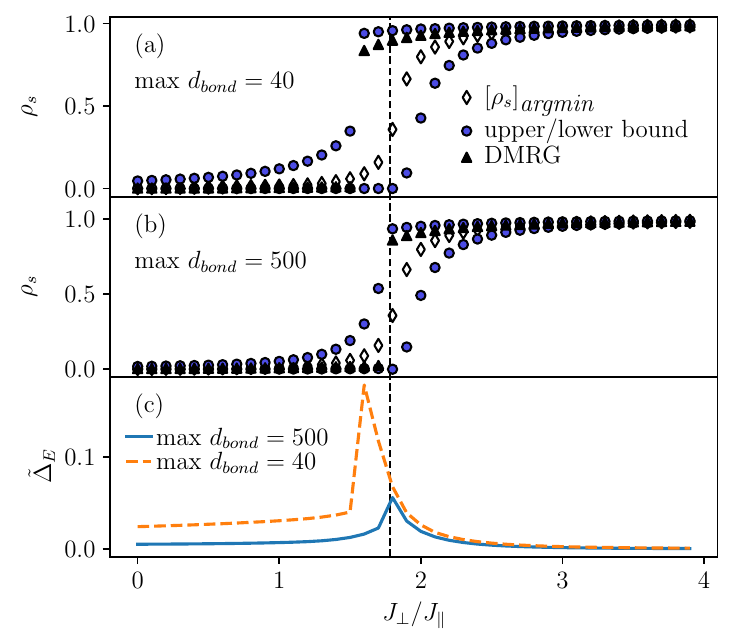}
    \caption{Properties of the FBH model for $L=4$ and $J_{x}/J_{\parallel}=0.6$. (a) Singlet density as a function of $J_{\perp}/J_{\parallel}$. The white diamonds are obtained from the moments in Eq.~\eqref{sdp-notation}, the blue circles are bounds obtained from Eq.~\eqref{sdp-obs-notation}, and the black triangles are obtained with DMRG with the maximum bond dimension $d_{bond}$ being 40. (b) Same as (a) but with the maximum bond dimension $d_{bond}$ being $500$. (c) Relative difference between the upper and lower bounds, see Eq.~\eqref{eq:bounddiff_E}. The dashed line corresponds to the transition point from Ref.~\cite{Stapmanns_18} at $J_{\perp}/J_{\parallel}=1.78$. }
    \label{fig:Jx06}
\end{figure}

Here, we investigate the influence of the variational upper bound on the results from Eq.~\eqref{sdp-obs-notation}. We follow Ref.~\cite{Stapmanns_18}, and study the dimer-singlet density, while incorporating the symmetries of the SDP:
\begin{equation}
\label{eq:rho}
    \rho_{s}=\frac{1}{4}(1-\sum\limits_{\alpha \in \{ x,y,z\}} \expval*{\hat \sigma^{\alpha}_{a,i,j}\hat \sigma^{\alpha}_{\bar{a},i,j}})= \frac{1}{4}(1-3 \expval{\hat \sigma^{x}_{a,i,j}\sigma^{x}_{\bar{a},i,j}}) \,.
\end{equation}  In Figs.~\ref{fig:Jx06}(a) and \ref{fig:Jx06}(b), we show $\rho_{s}$ for $L=4$ calculated with the density-matrix renormalization group (DMRG)~\cite{white_92} using a maximum bond dimension of 40 and 500, respectively. Furthermore, we show the value obtained by taking the \textit{argmin} of Eq.~\eqref{sdp-notation} and the upper and lower bounds on the singlet density; see Eq.~\eqref{sdp-obs-notation}. There, we observe that the \textit{argmin} correctly captures the formation of the singlet density but not the correct order of the transition. However, the usefulness of computing bounds becomes apparent. Whereas a poor variational ansatz (here, the bond dimension was intentionally chosen small for illustrative purposes) only provides a wrong estimate of the transition point, the SDP bounds give a regime in which the transition must take place. In (b), we see that as the upper bound is improved, this regime becomes narrower. In Fig.~\ref{fig:Jx06}(c), the relative energy differences using the different DMRG results are shown. The bounds calculated for $L=6$ are shown in Ref.~\cite{suppmat}. 

\begin{figure}
  \includegraphics[width=0.5\textwidth]{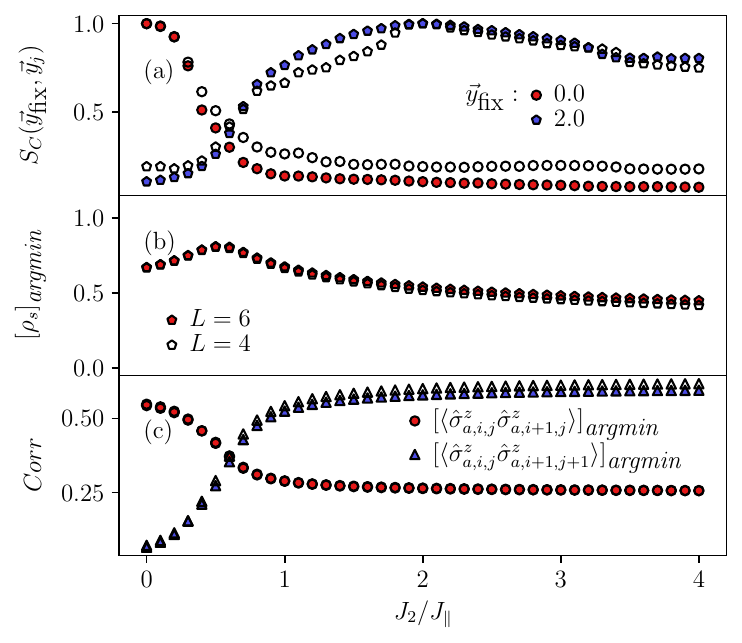}
    \caption{Different quantities calculated for the FBH-NNN model as a function of  $J_{2}/J_{\parallel}$. In all figures, the open symbols correspond to $L=4$ and the filled symbols to $L=6$. (a) Cosine similarity with
    $\vec{y}_{\textrm{fix}}$ at $(J_{\perp}/J_{\parallel},J_x/J_{\parallel} )=(1.5,0.1)$ with $J_{2}/J_{\parallel}=0.0$ (circles) and $J_{2}/J_{\parallel}=2.0$ (pentagons). (b) Singlet density obtained from the moments in Eq.~\eqref{sdp-notation}. (c) Nearest (circles) and next-nearest (triangles) neighbor  intralayer correlation functions also obtained from the moments in Eq.~\eqref{sdp-notation}.}
    \label{fig:multiple_phases}
\end{figure}

In Fig.~\ref{fig:multiple_phases}(a), we show $S_C(\vec{y}_{\textrm{fix}}, \vec{y}_j)$ for the FBH-NNN model for $\vec{y}_{\textrm{fix}}$ at $(J_{\perp}/J_{\parallel},J_x/J_{\parallel} )=(1.5,0.1)$ and different values of $J_{2}/J_{\parallel}$. We see that, when  $J_{2}/J_{\parallel}=0.0$ for $\vec{y}_{\textrm{fix}}$, $S_C(\vec{y}_{\textrm{fix}}, \vec{y}_j)$ decays rapidly as $J_{2}/J_{\parallel}$ is increased, signaling a phase transition. This is also reflected in $S_C(\vec{y}_{\textrm{fix}}, \vec{y}_j)$ when we set $J_{2}/J_{\parallel}=2.0$ for $\vec{y}_{\textrm{fix}}$, which is fairly constant for large $J_{2}/J_{\parallel}$ but rapidly decreases for small $J_{2}/J_{\parallel}$. The trend is further confirmed by comparing the system sizes $L=4$ and $L=6$, as $S_C(\vec{y}_{\textrm{fix}}, \vec{y}_j)$ decreases 
 to smaller values for the latter. This is also supported by plotting observables obtained by taking the \textit{argmin} from the moments in Eq.~\eqref{sdp-notation}. In Fig.~\ref{fig:multiple_phases}(b), we show the singlet density, which has a maximum around the transition region before slowly decaying for larger $J_{2}/J_{\parallel}$. We also plot the \textit{argmin} for the nearest and next-nearest neighbor intralayer correlation functions in Fig.~\ref{fig:multiple_phases}(c). The nearest neighbor correlations first decrease before going to a constant value. The next-nearest neighbor correlations increase rapidly with $J_{2}/J_{\parallel}$, indicating that the corresponding interaction becomes more and more important.  Note that further analysis of finite-size effects can be found in Ref.~\cite{suppmat}.

Finally, we demonstrate how simple manipulations of the SDP moments can provide us with useful information about the observables relevant to the transition. In Table~\ref{t1}, we show the moment that corresponds to the maximum value of $|\vec{y}_{\textrm{fix}}-\vec{y}_j|$. Whereas we do not have a physical interpretation for the moments for small $J_{\perp}/J_{\parallel}$, we see that as we are approaching the DTAF-DS transition line ($\approx 1.78$), the dimer correlations become more important before $\expval{\hat \sigma^x_{a,i,j}\hat \sigma^x_{\bar{a},i,j}}$ has the largest value. This moment is directly related to the singlet density, see Eq.~\eqref{eq:rho}, and thus, without any prior knowledge of the phases, we have extracted one observable relevant for the transition. 
\begin{table}
    \begin{tabular}{ | l | l | l |} 
    \hline
    $J_{\perp}/J_{\parallel}$ & Moment  \\ \hline 
    1.6 & $\expval{\hat \sigma^x_{a,i,j}\hat \sigma^x_{\bar{a},i,j+1}}$ \\ \hline
    1.7 & $\expval{\hat \sigma^x_{a,i,j}\sigma^x_{\bar{a},i,j}\hat \sigma^x_{a,i-2,j-1}\hat\sigma^x_{\bar{a},i-2,j-1}}$ \\ \hline
    1.8 &$\expval{\hat \sigma^x_{a,i,j}\hat \sigma^x_{\bar{a},i,j}\hat \sigma^x_{a,i-1,j-1}\hat \sigma^x_{\bar{a},i-1,j-1}}$ \\ \hline
    1.9 & $\expval{\hat \sigma^x_{a,i,j}\hat \sigma^x_{\bar{a},i,j}\hat \sigma^x_{a,i-1,j-1}\hat \sigma^x_{\bar{a},i-1,j-1}}$ \\ \hline
    2.0 & $\expval{\hat \sigma^x_{a,i,j}\hat \sigma^x_{\bar{a},i,j}}$ \\ \hline
    4.0 & $\expval{\hat \sigma^x_{a,i,j}\hat \sigma^x_{\bar{a},i,j}}$ \\ \hline
    \end{tabular}
    \caption{Moments corresponding to $argmax(|\vec{y}_{\textrm{fix}}-\vec{y}_j|)$ for the same parameters as in Fig.~\ref{fig:Jx06} but with $L=6$. Between $J_{\perp}/J_{\parallel}=2.0$ and $4$ and up to $J_{\perp}/J_{\parallel}=1.6$, the moment with the maximum value did not change. }
    \label{t1}
    \end{table}

\section{Cosine similarity and fidelity}
\label{cosimfid}
Here, we show that cosine similarity is an estimator of the fidelity between two pure states.
This motivates our choice of the metric, given the success of fidelity in the task of detecting phase transitions in previous works \cite{PhysRevE.74.031123}.

Consider the ground state $\ket{\psi}$ on a $L$ site chain for given Hamiltonian parameters. 
This has a Pauli decomposition:
$ 
\hat \rho=\ket{\psi}\bra{\psi}=\frac{1}{D}\sum_{\vec \alpha}
m_{\vec \alpha} {\hat P}_{\vec \alpha}\,,
$ 
where $D=2^L$, and ${\hat P}_{\vec \alpha}$ are defined in Eq.~\eqref{eq:P}
such that 
$ 
\tr{{\hat P}_{\vec \alpha} {\hat P}_{\vec \alpha'}}=D\, \delta_{\vec \alpha, \vec \alpha'}\,.
$ 
The moments $m_{\vec \alpha}$ fully describe the state $\hat \rho$.
If $\ket{\psi'}$ is the ground state for different Hamiltonian parameters (e.g., of another phase),  
$ 
\hat \rho'=\ket{\psi'}\bra{\psi'}=\frac{1}{D}\sum_{\vec \alpha}
m'_{\vec \alpha} {\hat P}_{\vec \alpha}\,,
$ 
the fidelity between the two pure states is given by 
\begin{eqnarray}
&& f=\tr{\rho \rho'}=D\frac{1}{D^2} \sum_{\vec \alpha}
m_{\vec \alpha} m'_{\vec \alpha}=\nonumber \\
&&
=D\,\mathbb{E} [m_{\vec \alpha} m'_{\vec \alpha}]
\xleftarrow{\text{estimates}} \frac{D}{K} \sum_{\vec \alpha \in {\mathcal{I}} } m_{\vec \alpha} m'_{\vec \alpha}\,,
\end{eqnarray} 
$\text{s.t.} \,\vert \mathcal{I} \vert =K$ \cite{estimator-comment},
 where $K$ is the number of moments considered (for our purposes $K \ll D^2$).
Defining $\vec{y}_i:=\{m_{\vec \alpha}\}_{\vec \alpha}$ and
$\vec{y}_j:=\{m'_{\vec \alpha}\}_{\vec \alpha}$ for $\vec\alpha \in {\mathcal{I}}$,
we obtain 
$f
\xleftarrow{\text{estimates}}
(D/K)\, \vec y_i \cdot \vec y_j$\,.
The rhs is an estimator of the fidelity.
Furthermore, because $\tr{\hat \rho ^2}=1$, we notice that 
\begin{eqnarray}
&& 1=\tr{\hat \rho \hat \rho}=\frac{D}{D^2} \sum_{\vec \alpha}
m_{\vec \alpha}^2 =
\nonumber
\\
&&=D\,\mathbb{E} [m_{\vec \alpha}^2 ]
\xleftarrow{\text{estimates}}
\frac{D}{K} \sum_{\vec \alpha \in {\mathcal{I}} } m_{\vec \alpha}^2=
\frac{D}{K} \Vert \vec{a} \Vert^2
\end{eqnarray}
\cite{estimator-comment}~(analogously for $\hat \rho'$).\\ 

Therefore, 
$K/D \xleftarrow{\text{estimates}} \Vert \vec{y}_i \Vert \Vert \vec{y}_j \Vert$\,.
Hence, the cosine similarity [see Eq.\,\eqref{eq:sc}] emerges as an estimator for the fidelity between the two ground states: 
\begin{equation}
   f  \xleftarrow{\text{estimates}}
   \frac{
   \vec y_i \cdot \vec y_j}{\Vert \vec{y}_i \Vert \Vert \vec{y}_j \Vert}\,.
   \label{estimator}
\end{equation}
These considerations have, for instance, some connections with \cite{flammia2011direct} but in a different context.

We remark that the estimate 
on the rhs of Eq.~\eqref{estimator}
is based on \textit{limited}-sample means and should be interpreted with caution [in the sense that, for our purposes, $K\sim poly(L)$ while $D\sim expo(L)$], despite being a valid estimator.
The estimator is {\it consistent}—that is, in principle it converges to the true value as 
$K\rightarrow D^2$---but it is not necessarily unbiased.  
Moreover, the moments used in Eq.~\eqref{eq:sc} are not the exact true moments, but rather estimates thereof.
Nevertheless, despite these limitations concerning the accuracy of the estimate in Eq.~\eqref{estimator} in the task of approximating the fidelity, our work shows that this estimator (cosine similarity) also performs effectively in detecting phase transitions.

 \bibliographystyle{biblev1}
 \bibliography{references}

\begin{thebibliography}{10}
\bibitem{estimator-comment}
\bibinfo{title}{Here, we use the sample mean as an estimator of the population
  mean, recognizing that it may be inaccurate due to the small sample size,
  biases etc.}
\end{thebibliography}

\end{document}